\documentclass[12pt, a4paper]{article}
\usepackage[affil-it]{authblk}

\newcommand{\avg}[1]{\langle #1 \rangle}

\newcommand{\R}[1]{\textrm{#1}}
\newcommand{\comment}[1]{}

\usepackage{graphicx}
\usepackage{cases}
\usepackage[breaklinks=true,colorlinks=true,linkcolor=blue,urlcolor=blue,citecolor=blue]{hyperref}
\usepackage[margin=1in]{geometry}

\begin{document}

\title{Optimal L\'{e}vy-flight foraging in a finite landscape}

\author[1]{Kun Zhao}
\author[1]{Raja Jurdak}
\author[1]{Jiajun Liu}
\author[2]{David Westcott}
\author[1]{Branislav Kusy}
\author[3]{Hazel Parry}
\author[1]{Philipp Sommer}
\author[2]{Adam McKeown}
\affil[1]{CSIRO Autonomous Systems, Brisbane, Queensland, Australia}
\affil[2]{CSIRO Land and Water, Atherton, Queensland, Australia}
\affil[3]{CSIRO Agriculture,  Brisbane, Queensland, Australia}

\maketitle

\begin{abstract}
We present a simple model to study L\'{e}vy-flight foraging in a finite landscape with countable targets. In our approach, foraging is a step-based exploratory random search process with a power-law step-size distribution $P(l) \propto l^{-\mu}$. We find that, when the termination is regulated by a finite number of steps $N$, the optimum value of $\mu$ that maximises the foraging efficiency can vary substantially in the interval $\mu \in (1,3)$, depending on the landscape features (landscape size and number of targets). We further demonstrate that subjective returning can be another significant factor that affects the foraging efficiency in such context. Our results suggest that L\'{e}vy-flight foraging may arise through an interaction between the environmental context and the termination of exploitation, and particularly that the number of steps can play an important role in this scenario which is overlooked by most previous work. Our study not only provides a new perspective on L\'{e}vy-flight foraging, but also opens new avenues for investigating the interaction between foraging dynamics and environment as well as offers a realistic framework for analysing animal movement patterns from empirical data. \\


\end{abstract}

\section{Introduction} 

Understanding animal movement is crucial for understanding ecological and evolutionary processes in nature and has a wide range of applications such as ecosystem management, species conservation and disease control \cite{smouse2010stochastic, bartumeus2007levy, ortiz2006use, caro1999behaviour, bousquet2004multi}. 

In the last decade it has been widely observed that the movements of many animal species, from albatrosses \cite{viswanathan1996levy, humphries2012foraging} to spider monkeys~\cite{ramos2004levy}, honey bees \cite{reynolds2007displaced} to deer \cite{focardi2009adaptive},  and marine predators \cite{sims2008scaling} to human foragers~\cite{hills2013adaptive, raichlen2014evidence}, appear to exhibit L\'{e}vy-flight patterns, i.e. the step-size distribution can be approximated by a power-law $P(l) \propto l^{-\mu}$ with $1<\mu \le 3$. Despite this apparent similarity, there is an ongoing debate in the scientific community over the existence of L\'{e}vy-flights in animal movement and the methodology of verifying L\'{e}vy-flights from empirical data \cite{benhamou2007many, edwards2007revisiting, edwards2011overturning}. In the meanwhile, scientists, especially theorists, are keen on a question from a theoretical perspective: if the existence of L\'{e}vy-flight in animal movement is true,  why do animals perform L\'{e}vy-flight? This question fascinates researchers from various disciplines from ecology to physics \cite{viswanathan2008levy, viswanathan2011physics, james2011assessing, bartumeus2009optimal, viswanathan2000levy, bartumeus2013stochastic}. 

One common approach to the origin of animal movement patterns is to use the scheme of optimizing random search \cite{bartumeus2005animal, da2009random, reynolds2009levy}. In a random search model, single or multiple individuals search a landscape to locate targets whose locations are not known \textit{a priori}, which is usually adopted to describe the scenario of animals foraging for food, prey or resources. The locomotion of the individual has a certain degree of freedom which is characterised by a specific search strategy such as a type of random walk, and is also subject to other external or internal constraints, such as the environmental context of the landscape or the physical and psychological conditions of the individual. It is assumed that a strategy that optimises the search efficiency can evolve in response to these constraints, and the movement is a consequence of the optimisation on random search.  

A seminal work by Viswanathan \textit{et al.}  \cite{viswanathan1999optimizing} first studied L\'{e}vy-flight foraging through the scheme of optimizing random search. In their model, a forager searches for targets using a random walk with the aforementioned power-law step-size distribution. The forager will keep moving until a target is 'encountered', i.e. a target lies within its limited perception range . The search efficiency is defined as the encounter rate of targets, namely the number of visited targets per unit moving distance. The model considers two scenarios: (i) non-destructive foraging, in which the targets are revisitable, and (ii) destructive foraging, in which the targets are depleted once visited. For non-destructive foraging, when the targets are sparsely distributed, there exists an optimum exponent $\mu_{opt} \approx 2$ that maximises the search efficiency, corresponding to the L\'{e}vy-flight strategy. For destructive foraging, $\mu_{opt} \rightarrow 1$, corresponding to the ballistic motion, is the optimum solution. It is worth noting that this model only captures an idealised scenario in which learning and prey-predator relationships are ignored \cite{viswanathan1999optimizing}, and other random search strategies such as intermittent random search can outperform the L\'{e}vy-flight strategy in some scenario \cite{benichou2007minimal}. 

Recent studies also turn attention to the substantial difference of the value of $\mu_{opt}$ observed in empirical data. In fact the value of $\mu_{opt}$ among different species can range from $\mu_{opt} \approx 1.59$ for human beings \cite{gonzalez2008understanding} to $\mu_{opt} \approx 2.4$ for bigeye tuna \cite{sims2008scaling}. $\mu_{opt}$ can also varies significantly among individuals within the same species, e.g. the value of $\mu_{opt}$ can go from $1.18$ to $2.9$ in jellyfish \cite{hays2012high}. It has been shown that intermediate values of the optimal exponent $1 < \mu_{opt} \le 2$ can emerge in the crossover regime between non-destructive and destructive foraging in which targets are regenerated after a period $\tau$ once depleted \cite{santos2004optimal}, or arise in response to the landscape heterogeneity \cite{raposo2011landscape}. A recent study by Palyulin \textit{et al.} \cite{palyulin2014levy} shows that, in the case of searching for a single target in a infinite one-dimensional space, when an external drift (underwater current, atmospheric wind and etc.) is present, $\mu_{opt}$ can also vary in the interval $2 \le \mu_{opt} \le 3$. 

In this paper, we propose a simple model to study L\'{e}vy flight foraging in a finite two-dimensional landscape with countable targets. In our model, we consider foraging to be a step-based exploratory search process for distinct targets subject to termination. The forager can revisit the targets and the foraging efficiency is defined as the total number of discovered targets per unit moving distance in this process. We find that, different optimal search strategies can emerge in adaption to different termination conditions. In particular, when the termination is regulated by a finite number of steps $N$, optimum L\'{e}vy-flight strategies with various exponent $\mu_{opt}$ can emerge due to the exploitation of the landscape. In this case, the value of $N$, along with the landscape features such as the landscape size and the number of targets, can play an important role in determining the value of $\mu_{opt}$.  When the termination is regulated by a finite moving distance $\mathcal{L}$, the best strategy is always ballistic motion corresponding to $\mu_{opt} \rightarrow 1$. To capture more complex foraging dynamics, we also consider subjective returning in our model through an exploration-return mechanism \cite{song2010modelling}, and demonstrate that this returning can be another factor that affects the foraging efficiency. Our study not only sheds new light on the the understanding of L\'{e}vy-flight foraging, but also provides a new modelling framework to study animal movement patterns.

\section{Model}

The foraging takes place in a finite two-dimensional $L \times L$ squared landscape with periodic boundary conditions (when the forager moves across the boundary it will continue the movement on the opposite side of the landscape). There are $K$ targets distributed uniformly over the landscape, corresponding to a density $\rho = L^2 / K$.  The forager can detect a target within its perception range $r_v$.  The mean free path of the system $\lambda$ is therefore given by $\lambda = (2\rho r_v) ^ {-1}$, which indicates the average straight-line moving distance of detecting or `encountering' a target in the landscape. Without loss of generality we set $r_v = 1$, so $\lambda = (1/2\rho)$. In this paper, we assume the targets are revisitable, analogous to the case of non-destructive foraging \cite{viswanathan1999optimizing}. The goal of the forager is to explore the landscape to find new targets.  

\begin{figure}[htb!]
\centering
\includegraphics[width=6in]{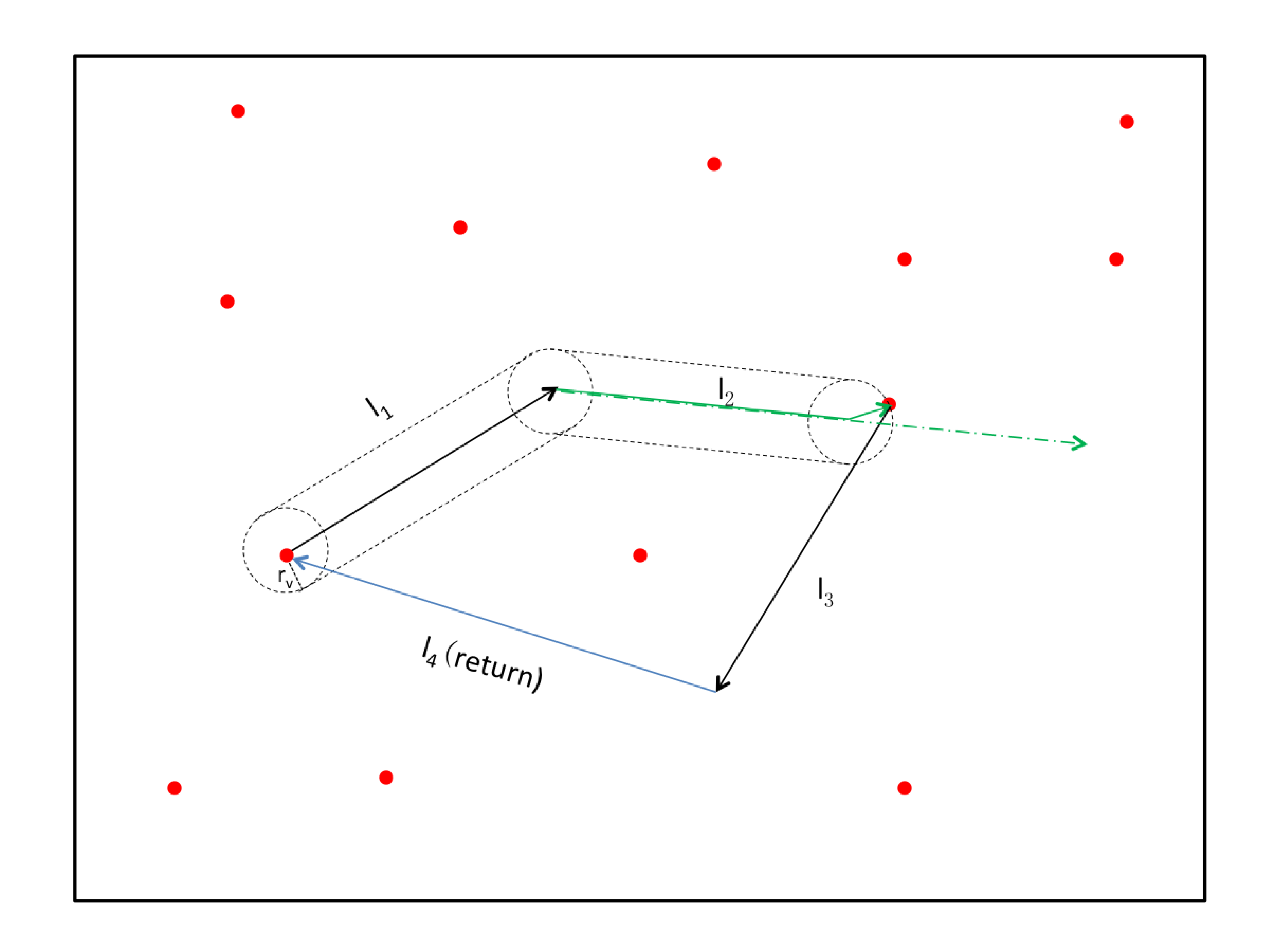}
\caption{A schematic diagram of the model. This diagram shows a foraging process with $N=4$ steps. The red dots represent targets. The cylinder formed by black dashed boundaries indicates the detection area during an exploration step. In step one, the forager leaves a target and detects no new target during this step. Note that the forager ignores the departure target. In step two, the forager decides to do exploration and detects a target during this step. Therefore the original probabilistic step (the green dashed-dotted line) is truncated to a shorter actual step $l_2$ (the green solid line). Step three is similar to step one. In step four the forager decides to return and it flies straight back to the departure target in step one.  In a return step, the forager is not attempting to detect targets.  }
\label{diagram}
\end{figure}

The foraging process is a step-based stochastic process with an exploration-return mechanism.  At each step $n$, the forager first decides the type of the step movement: exploration or return. Let us denote the probability of choosing exploration by $p_n$ and the probability of choosing return by $q_n = 1 - p_n$. Moreover, we use $S_n$ to denote the number of distinct targets discovered by the forager up to step $n$, $\mathcal{L}_f$ to denote the accumulated moving distance since the forager leaves its last visited target, $l_n$ to denote the moving distance of step $n$, and $\mathcal{L}_n = \sum l_n$ to denote the total moving distance up to step $n$. The foraging movement at step $n$ is performed as follows (illustrated in Figure \ref{diagram}): 
\begin{enumerate}
\item If the decision is exploration,  the forager will perform random search in this step. The step-size $l$ and the turning angle $\theta$ are drawn randomly from the pre-defined distribution functions $P(l)$ and $P(\theta)$.  During the step movement, the forager will continuously detect targets. If a target is detected during the step movement, the forager will move to the target in a straight line and the step movement will be truncated. The actual moving distance $l_n = l - \Delta l$ in this case is smaller than the probabilisic moving distance $l$ and we set $\mathcal{L}_f \rightarrow 0$. There are two situations of detecting a target: (a) The target is a new target that has not been discovered before. Then we update $S_n$ by $S_n \rightarrow S_n + 1$ and the location of this new target is memorised by the forager. (b) The target is a previously visited target. In this case we do not update $S_n$. One should note that if this step is the first step the forager leaves a target, the forager will ignore that target in detection to avoid trapping.  If no target is detected in this step, we update $\mathcal{L}_f$ by $\mathcal{L}_f \rightarrow \mathcal{L}_f + l$.
\item If the decision is returning,  the forager will move to one of the previously visited targets in a straight line. Note that the forager does not attempt to detect targets in a return step, which is analogous to the `blind' phase in intermittent random search \cite{benichou2007minimal}. We assume that the forager can memorise the locations of all previously visited targets and randomly decide on the target of the return phase. In this initial model we focus on this simple approach to modelling memory and leave more complicated memory process to future work. 
\end{enumerate}

Here we assume that the foraging process starts at a random target in the landscape. That target can be understood as the base of the forager, and its location is recorded in the initial memory of the forager. Therefore the forager can have at least one location to choose in the return phase. We use an indicator function $\Theta_n \in \{0,1\}$ to characterise the termination condition of the process. When a step is performed, we update $n$ by $n \rightarrow n+1$ and check the value of  $\Theta_n$.  The process will be continued if $\Theta_{n} = 0$, and be terminated once $\Theta_{n=N} = 1$ where $N$ denotes the total number of steps upon termination. 

We then define the search efficiency $\eta$ as the ratio of the total number of distinct targets discovered by the forager to the total moving distance upon termination, which yields
\begin{equation}
\eta \equiv \frac{S_N}{\mathcal{L}_N}.
\label{eta} 
\end{equation}
One should note that, besides the subjective returning in the return phase, the forager can also revisit a previously discovered target if it lies within the forager's perception range in the exploration phase. The revisitation during exploration may occur in two scenario: (1) the forager taking advantage of the chance proximity of a previously discovered target to relieve movement constraints (e.g. to rest or supplement energy) prior to reinitiating exploration; (2) the forager can detect a target within its perception range, but is not able to identify the resource availability in the detected target without a revisitation. In such context, the target can be understood to be a site that contains resource (e.g. a tree with fruits). The random search dynamics in the exploration phase in our model resembles the non-destructive foraging in \cite{viswanathan1999optimizing}, while the definition of search efficiency that only accounts for distinct targets discovered by the forager resembles the case of destructive foraging with exploitation.

Finally, we specify the random search dynamics and the exploration-return mechanism. In this paper, we use a power-law step-size distribution for the random search, which yields
\begin{equation}
P(l) = \frac{\mu - 1}{l_0}\left( \frac{l}{l_0}\right) ^ {-\mu},  ~~ l \ge l_0 
\label{Pl}
\end{equation}
where $1 < \mu \le 3$ is the power-law exponent which serves as a control parameter of the random search foraging strategy. The lower bound $l_0$ represents the natural limit of step-size. The movement converges to the Gaussian (Brownian motion) when $\mu \ge 3$, and to  ballistic motion when $\mu \rightarrow 1$. For simplicity in this paper we set $l_0 = r_v = 1$ and we use a uniform distribution $P(\theta) = 1/ 2 \pi$ for the turning angle.  
 
To specify $p_n$ and $q_n$, we assume that the forager makes a decision between exploration and return according the following rule: \textit{the longer it has traveled since it leaves the last visited target, the more likely it will decide to return}. The decision rule here reflects the accumulated resistance on the forager's will for exploration as the moving distance increases,  such as the decline of energy level, accumulated stress of finding no new targets, or other behavioural regularities. Therefore $p_n$ can be represented by a non-increasing function of $\mathcal{L}_f$. Specifically in this paper we define $p_n$ as an exponential function

\begin{equation}
p_n = \exp(- \beta\mathcal{L}_f ) 
\end{equation}
where $\beta$ is a control parameter for tuning the intensity of such returning. A higher $\beta$ leads to a higher likelihood of subjective returning during exploration, which models more conservative foraging behaviour. Similar dynamics have been observed empirically in both animals \cite{niv2002evolution} and human \cite{zhao2011social}. The subjective returning can be used to model the home-return patterns \cite{song2010modelling} or the central-place foraging dynamics \cite{reynolds2008optimal, steingrimsson2008multiple}. When $\beta = 0$ such that $p_n = 1$, the model is without an  return phase and the foraging dynamics resembles the non-destructive foraging in \cite{viswanathan1999optimizing}. 

In summary, the foraging in our model is tuned by five parameters, namely $\mu$, $N$, $L$, $K$, and $\beta$. The power-law exponent $\mu$ characterises the random search strategy. The landscape size $L$ and the number of targets $K$ represent the environmental features. The number of steps $N$ regulated by the indicator function $\Theta_n$ specifies the termination of the foraging. The intensity of subjective returning $\beta$ is an additional parameter to characterise more complex foraging dynamics.

\section{Results and Discussions}
\subsection{The case with $\beta=0$}
We first discuss the simple case without the subjective returning by setting $\beta = 0$ and $p_n = 1$. To gain better insight into the model, we derive a closed-form expression for the search efficiency $\eta$ using mean-field approximation. We then investigate the relationship between the optimal exponent $\mu_{opt}$ and other parameters of the model using a mixture approach of numerical simulations and analytical evaluation.
 
\subsubsection{Approximate mean-field solution} 
\begin{figure}[htb!]
\centering
\vspace{1in}
\includegraphics[width=5.3in]{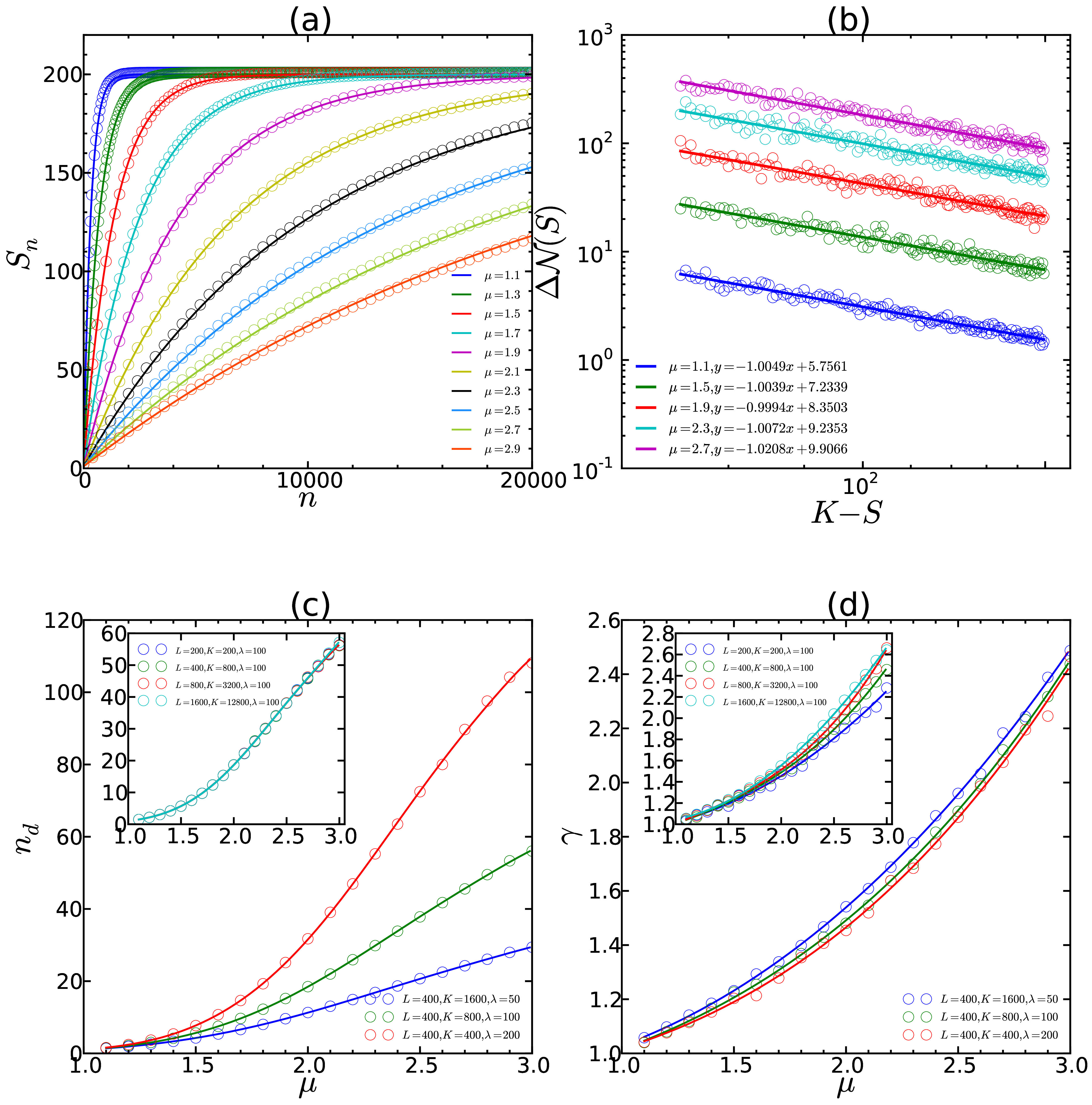}
\caption{(a) The number of discovered targets $S_n$ versus step number $n$.  The curves in different colours from top to bottom correspond to various $\mu \in [1.1, 2.5]$ with an interval of $0.2$.  Here we use the landscape size $L = 200$ and the number of patches $K = 200$. The dots represent simulation results averaged for 100 realisations and the solid lines represent the corresponding mean-field solution given by Eq.(\ref{Sn}). (b) The mean number of steps between the discovery of the $S^{\textit{th}}$ target and the $(S+1)^{\textit{th}}$ target $\Delta\mathcal{N}(S)$ versus the number of undiscovered targets $K - S$. The dots represent simulation results and the lines represent the corresponding linear regression fit. (c) The mean number of steps between two consecutive detections $n_d$ as a function of $\mu$ for different values of $L$ and $K$. The solid lines represent the nonlinear fitting $n_d(\mu) \sim A(\mu) \lambda^{\frac{\mu-1}{2}}$. (d) The constant coefficient $\gamma$ as a function of $\mu$ for different values of $L$ and $K$. The solid lines represent the cubic polynomial fitting. }
\label{fig4}
\end{figure}

To evaluate Eq.(\ref{eta}), we first calculate the numerator $S_N$. We denote by $\mathcal{N}(S)$ the mean number of steps from the beginning to the discovery of the $S^{\textit{th}}$ new target, and $\Delta \mathcal{N}(S) \equiv \mathcal{N}(S+1) - \mathcal{N}(S)$ the mean number of steps between the discovery of the $S^{\textit{th}}$ target and the $(S+1)^{\textit{th}}$ target. Intuitively, since the foraging is confined in a finite landscape with limited number of targets, as the forager explores the landscape more,  the number of undiscovered targets decreases and discovering new targets becomes more difficult. If the forager detects a target at step $n$, the probability that the detected target is new, denoted by $p_{new}$,  should be approximately proportional to the number of undiscovered targets at step $n$, namely $p_{new} = (K - S_n)/\gamma K$ where $\gamma$ is a constant coefficient. Therefore, to discover one more new target, the forager has to make $1/p_{new}$ detections on average, which gives
\begin{equation}
\Delta \mathcal{N} \approx \frac{n_d}{p_{new}} = \frac{K}{K-S_n} \gamma n_d
\label{dN}
\end{equation}
where $n_d$ is the mean number of steps between two consecutive detections. The assumption that $\Delta \mathcal{N} \propto (K - S)^{-1}$ indicated by Eq.(\ref{dN}) is supported by the numerical simulation, as shown in Figure \ref{fig4}(b). The increment of $S_n$ at each step $n$ can be written as:
\begin{equation}
\frac{dS_n}{dn} \approx \frac{S_{n+1} - S_n}{(n+1)-n}=\frac{1}{\Delta \mathcal{N}(S_n)} \approx \frac{K - S_n}{K}\frac{1}{\gamma n_d}.
\label{ds_eq}
\end{equation}
The differential equation Eq.(\ref{ds_eq}) can be solved with initial condition $S_0 = 1$, which yields 
\begin{equation}
S_n \approx K-(K-1)e^{-\frac{n}{K \gamma n_d}}.
\label{Sn}
\end{equation}

Then we calculate the denominator $\mathcal{L}_N$. In mean-field approximation, $\mathcal{L}_N$ can be simply expressed as $\mathcal{L}_N \approx N \avg{l}$ where $\avg{l}$ is the mean step-size given by \cite{viswanathan1999optimizing}
\begin{eqnarray}
\avg{l} &\approx& \int_{l_0}^{\lambda}lP(l)dl + \int_{\lambda}^{\infty} \lambda P(l)dl \nonumber \\
&=& \left( \frac{\mu-1}{2-\mu} \right) \left( \frac{\lambda^{2-\mu} - l_0^{2-\mu}}{l_0^{1-\mu}} \right) + \frac{{\lambda}^{2 - \mu}}{l_0^{1-\mu}}.
\label{avg_l}
\end{eqnarray}
Via Eq.(\ref{Sn}) and Eq.(\ref{avg_l}) we are able to calculate the search efficiency $\eta$. 

To evaluate $S_n$ analytically through Eq.(\ref{Sn}) we still need to know $n_d$ and $\gamma$. Unfortunately obtaining an exact closed-form expression for $n_d$ and $\gamma$ is still a difficult mathematical issue to date. Therefore we use curve fitting to obtain the value of $C = \gamma n_d$ in Eq.(\ref{Sn}) from simulation. The advantage of this approach is that both $\gamma$ and $n_d$ remain approximately constant as $n$ increases. Therefore we can obtain the value of $C$ by performing the simulation up to a small number of steps (e.g. up to $20\%$ targets in the landscape have been discovered in our study), and then use the analytical solution to extrapolate the results for large $n$. $S_n$ evaluated by Eq.(\ref{Sn}) with a fitting parameter $C$ is in good agreement with the numerical simulation, as shown in Figure \ref{fig4}(a). With $L$ and $K$ given, to obtain a continuous curve for $\eta(\mu)$ by performing numerical simulations for a small number of discrete values of $\mu$, we can also use curve fitting to find $\gamma(\mu)$ and $n_d(\mu)$, as shown in Figure \ref{fig4} (c) - (d). In particular we find that $\gamma(\mu)$ can be well approximated by a cubic polynomial function, and $n_d(\mu)$ can be well approximated by $n_d(\mu) \sim A(\mu) \lambda^{\frac{\mu-1}{2}}$ as suggested by \cite{viswanathan1999optimizing, buldyrev2001average} where $A(\mu)$ can be fitted by a fifth degree polynomial function. Our results also suggest that,  with $\mu$ given, $\gamma$ depends on both $L$ and $K$ while $n_d$ only depends on $\lambda$, as shown in Figure \ref{fig4} (c) - (d).

We note that both the denominator and numerator in Eq.(\ref{eta}), denoting the total moving distance $\mathcal{L}_N = N \avg{l}$ and total number of discovered targets $S_N$, respectively, are a decreasing function of $\mu$. Intuitively this is easy to understand. In an N-step truncated L\'{e}vy flight, decreasing $\mu$ not only leads to a larger mean step-size as indicated by Eq.(\ref{avg_l}), but also enlarges the searched area so that more new targets can be discovered by the forager. However their rates of increase with respective to the decrease of $\mu$ differ, which may lead to an optimal value for $\mu$ that maximises their ratio. In the following we study how the optimal exponent $\mu$ depends on the number of steps $N$, the landscape size $L$ and the number of patches $K$. 

\subsubsection{Number of steps $N$}

\begin{figure}[htb!]
\centering
\vspace{1in}
\includegraphics[width=6in]{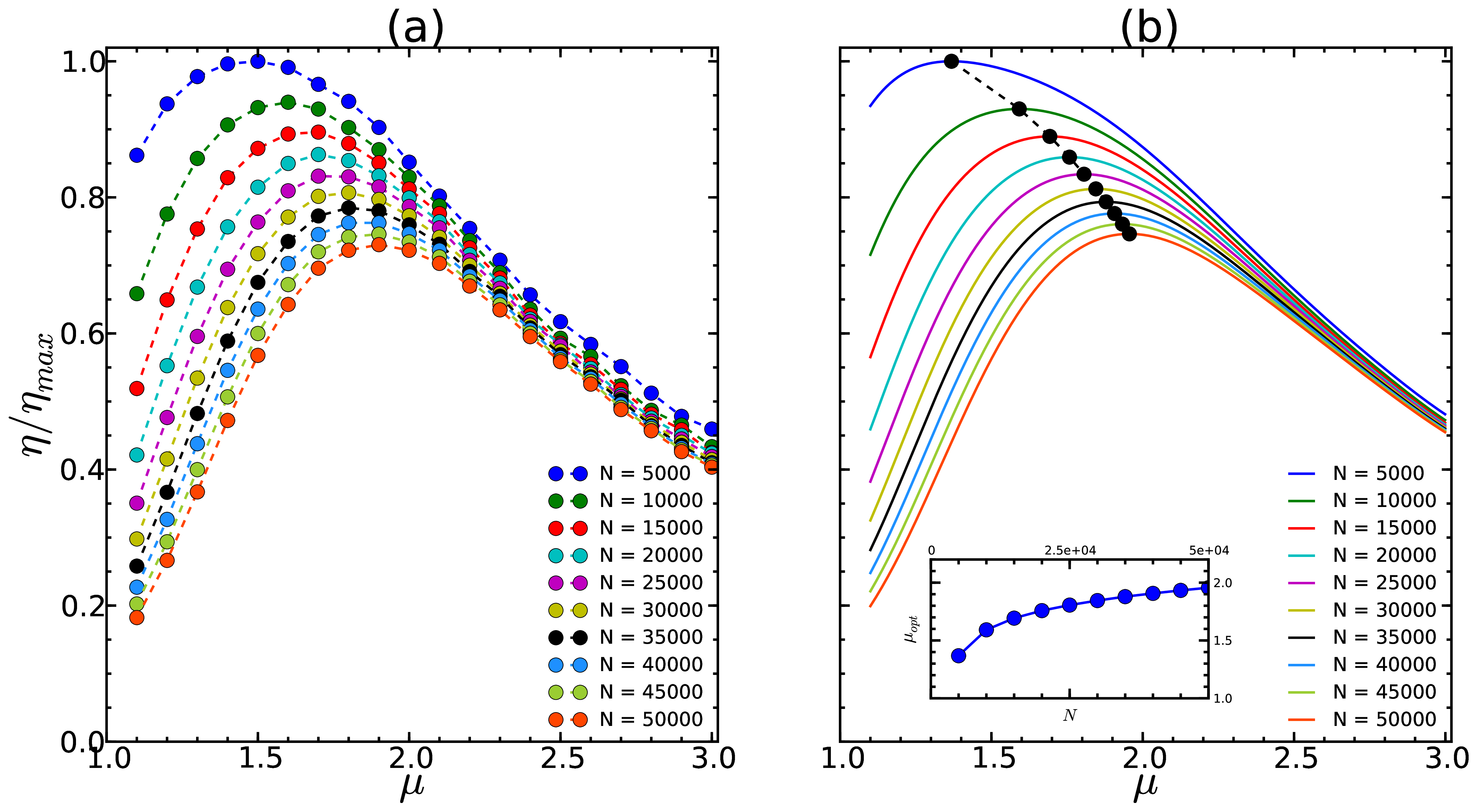}
\caption{(a) The rescaled search efficiency $\eta / \eta_{max}$ versus the power-law exponent $\mu$ for different values of the total number of steps $N$ from numerical simulation with the intensity of subjective returning $\beta = 0$, the landscape size $L = 1000$, the number of patches $K = 5000$ and the termination condition $\Theta_n = \delta(n - N)$.  The results are averaged over $100$ realizations. The rescaling factor $\eta_{max} = \eta(\mu=\mu_{opt}, N = 5000)$ is the overall maximum search efficiency. (b) $\eta / \eta_{max}$ versus $\mu$ from mean-field calculation.  The black dots indicate the peaks of the curves. The inset of panel (b) shows $\mu_{opt}$ versus $N$.  The curves in panel (a) and (b) with different colours from top to bottom correspond to various $N\in [5\times10^3,5\times10^4]$ with an interval of $5000$.  }
\label{fig3}
\end{figure}

To study the influence of the number of steps $N$, we use an indictor function $\Theta_n = \delta(n - N)$ such that the foraging process is terminated at $n = N$ step for any given value of $\mu$. Here $\delta(x)$ is the Kronecker delta function such that $\delta(x) = 0$ if $x =0$ and $\delta(x) = 1$ if $x = 0$. We perform numerical simulation with  $L = 10^3$, $K = 5000$, $\lambda = 100$. The search efficiency $\eta$ is then fully determined by the number of foraging steps $N$ and the power-law exponent $\mu$, which yields $\eta=\eta(\mu, N)$. We observe that for a given value of $N$, there exists an optimal exponent $\mu_{opt}$ which maximises $\eta$. Surprisingly we find that the value of $u_{opt}$ shifts substantially as $N$ varies and is overall an increasing function of $N$, as shown in Figure \ref{fig3}(a).

 A simple explanation for the presence of such optimality relates to the increased difficulty in discovering new targets later in the foraging process, particularly for smaller $\mu$. Recall that the forager can discover new targets more rapidly in the beginning (high-efficient stage) but then enters into a difficult stage for discovering new targets in the latter steps (low-efficient stage) as the number of new targets drops. The random search with smaller $\mu$ will enter this low-efficient stage earlier in the foraging process, as shown in Figure~\ref{fig4}(a). The gain in $S_N$ from reducing $\mu$ will decline steadily, and will eventually no longer compensate for increases in $\mathcal{L}_N$; at that point, $\eta$ reaches its maximum.

The optimal exponent $\mu_{opt}$ as a function of $N$ can be obtained by solving $\frac{\partial \eta(\mu, N)}{\partial \mu} \bigg |_{\mu = \mu_{opt}} = 0$, as shown in the inset of Figure \ref{fig3}(b). We note that when $N \rightarrow 1$ the optimal exponent will approach $\mu_{opt} \rightarrow 1$ approximately corresponding to ballistic motion. The optimal exponent can also approach $\mu = 3$ for a sufficiently large $N$, which means the optimal random search can go through a transition from L\'{e}vy-flight ($1 < \mu < 3$) to Brownian motion ($\mu \ge 3$) as $N$ increases.

\subsubsection{Landscape size $L$ and number of patches $K$}
\begin{figure}[htb!]
\centering
\vspace{1in}
\includegraphics[width=6in]{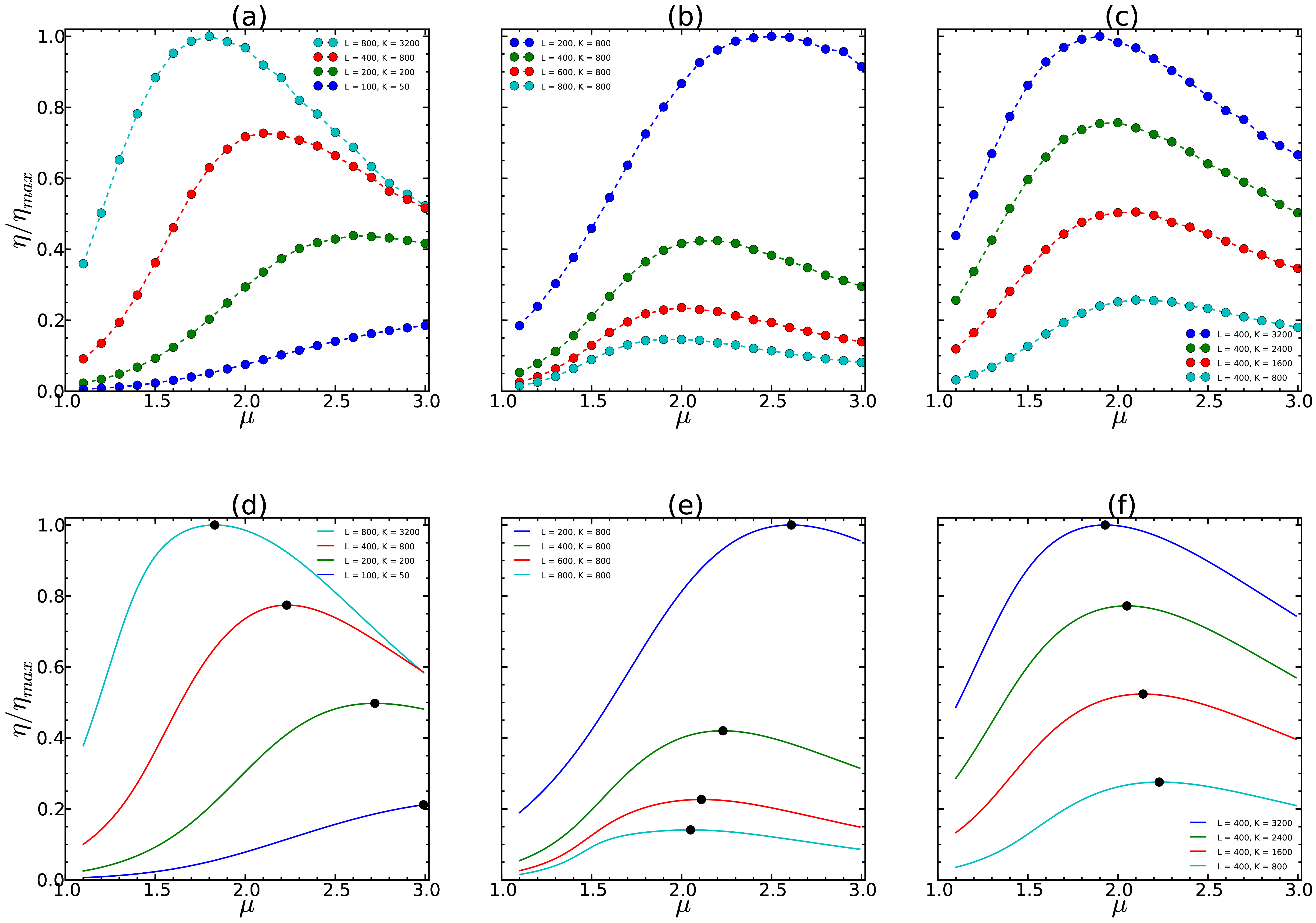}
\caption{ Panel (a) - (c) show the rescaled search efficiency $\eta / \eta_{max}$ versus the power-law exponent $\mu$ for different values of the landscape size $L$ and the number of patches $K$ from numerical simulation with the intensity of subjective returning $\beta = 0$ and the termination condition $\Theta_n = \delta(n - 50000)$. The results are averaged over $100$ realizations. The curves in panel (a) from bottom to top correspond to $(L,K) = (100,50), (200,200), (400, 800), (800, 3200)$  such that the mean free path $\lambda =100$ remains constant.  The curves in panel (b) from top to bottom correspond to $L = 200,400,600,800$ with $K=800$ kept constant .  The curves in panel (c) from top to bottom correspond to $K = 800, 1600, 2400, 3200$ with $L=400$ kept constant. Panel (d) - (f) show $\eta / \eta_{max}$ evaluated by the mean-field solution, corresponding to the results from numerical simulation in Panel (a) - (c), respectively. The black dots indicate the peak of the curves.}
\label{fig8}
\end{figure}

We then study how the foraging efficiency depends on the landscape size $L$ and number of patches $K$ under the termination condition $\Theta_n = \delta(n - N)$. We evaluate the search efficiency by varying the landscape size $L$ and the number of patches $K$ in three different combinations, as shown in Figure \ref{fig8}: (1) varying $L$ and $K$ simultaneously with the mean free path $\lambda = L^2 / 2K$ kept constant; (2) varying $L$ with $K$ kept constant; (3) varying $K$ with $L$ kept constant. 

Intuitively,  with $N$ and $\lambda$ given, as $K$ increases, the optimal exponent $\mu_{opt}$ will shift to a smaller value since the foraging can stay in the high-efficient stage for longer total moving distance, accounting for the situation in combination (1). Similarly, with $N$ and $K$ given, as $\lambda$ decreases, $\mu_{opt}$ will shift to a larger value, since the average distance for discovering new targets becomes shorter and the low-efficient stage comes earlier for smaller $\mu$, accounting for the situation in combination (2). The couple effect of increasing $K$ and decreasing $\lambda$ is studied through combination (3). In this case, as $K$ increases, $\mu_{opt}$ still shifts to a smaller value, but the shifting is not as significant as the case in combination (1).  

Our results suggest that the foraging efficiency and the optimal foraging strategy characterised by $\mu_{opt}$ can have an interesting dependence on the landscape features under specific termination condition.

\subsubsection{Termination condition $\Theta_n$}
The foraging defined in our model is considered to be an individual search process in a finite area subject to termination. This setting can account for two scenarios in the real-world: (1) foraging within a restricted area during a certain foraging season; (2) foraging in a fractal landscape (where resources are clustered in patches) such that searching each patchy area can be viewed as a single process. In scenario (2), due to the depletion of new targets, the forager has to decide  on when to terminate the foraging. The indicator function $\Theta_n$, which characterises the termination condition and regulates the number of steps $N$, can reflect the animal's prior experience, its physical conditions or behavioural regularity, or other environmental conditions such as seasonal variation that can has an influence on the forager's decision on changing foraging area. In this context, various optimal strategies may evolve in response to different termination conditions coupled with the environmental context of the foraging area. 

\begin{figure}[htb!]
\centering
\vspace{1in}
\includegraphics[width=6in]{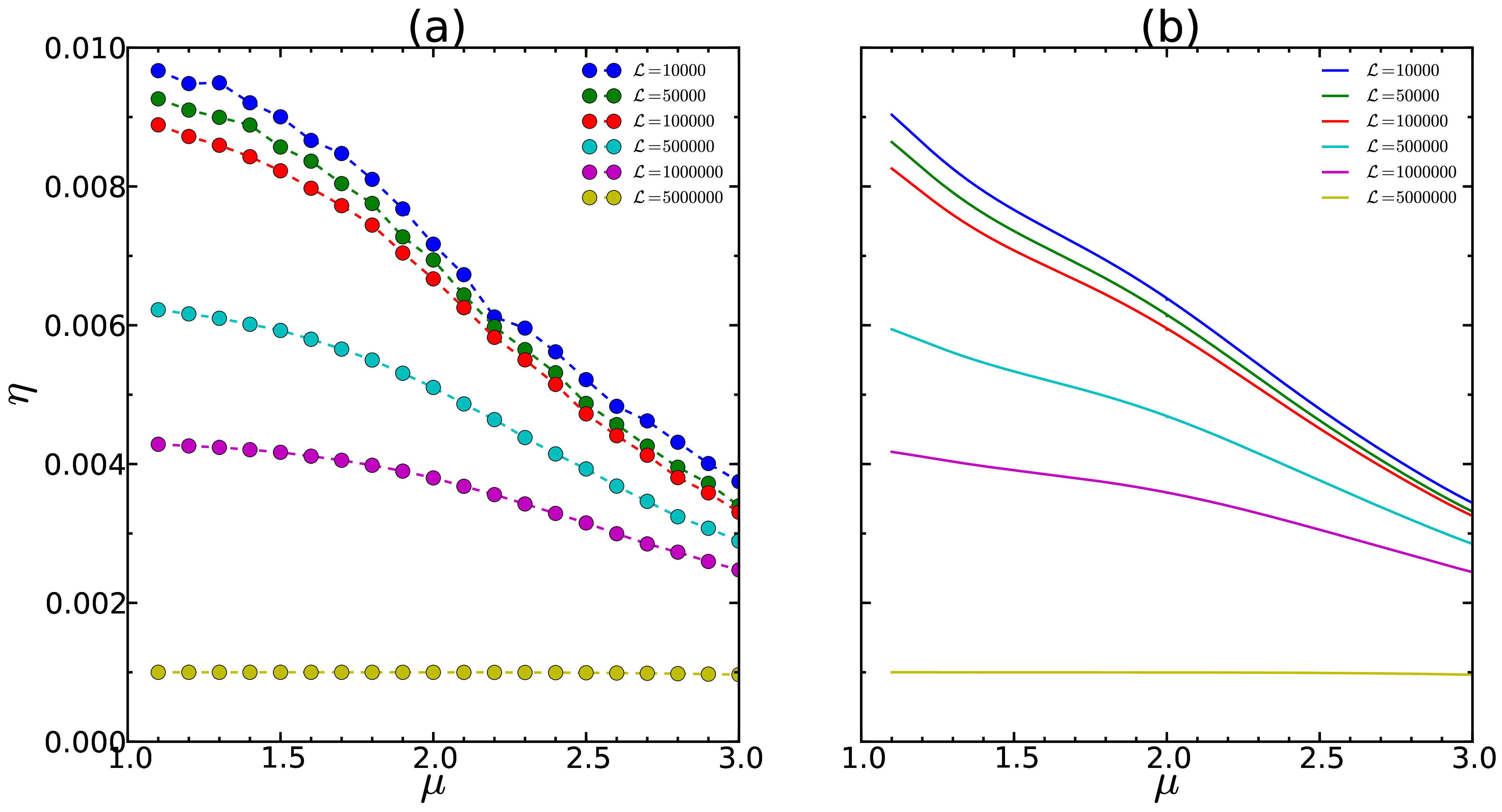}
\caption{ (a) The search efficiency $\eta$ versus the power-law exponent $\mu$ for different values of the threshold moving distance $\mathcal{L}$ from numerical simulation with the intensity of subjective returning $\beta = 0$, the landscape size $L = 1000$, the number of patches $K = 5000$ and the termination condition $\Theta_n = \theta(\mathcal{L}_f - \mathcal{L})$ where $\mathcal{L}_f$ is the accumulated moving distance. The results are averaged over $100$ realizations. (b) $\eta$ evaluated by the mean-field solution by setting the total number of steps $N$ by $N \approx \mathcal{L} / \avg{l}$ in Eq.(\ref{Sn}). }
\label{fig-Lmax}
\end{figure}

For comparison, we study another termination condition $\Theta_n = \theta(\mathcal{L}_f - \mathcal{L})$, where $\theta(x)$ is the step function such that $\theta(x) = 0$ if $x < 0$ and $\theta(x) = 1$ if $x \ge 0$. In this case, the foraging is terminated once the the accumulated moving distance $\mathcal{L}_f$ exceeds a threshold $\mathcal{L}$, and the best strategy is always ballistic motion corresponding to $\mu_{opt} \rightarrow 1$, as shown in Figure \ref{fig-Lmax}. The results here implies that, if the forager always prefers exploring the landscape with a prefixed length of moving distance $\mathcal{L}$ (or a certain amount of energy for exploration if we consider energy expenditure is proportional to $\mathcal{L}$), the optimal search strategy will evolve to ballistic motion regardless of the choice of $\mathcal{L}$. On the other hand, if the forager explores the landscape with a random moving distance $\mathcal{L}_N$ (or a random amount of energy) regulated by a certain number of steps $N$, various optimal search strategies from ballistic motion to L\'{e}vy-flight to Brownian motion can emerge depending on the landscape features. This situation is more likely to occur in scenario (1), in which the animal may subject to daily regularity during the foraging season. For instance, the animal may only perform a regular number of foraging steps (trips/bouts) per day, and the moving distance in each step is randomly distributed. It is also interesting to note that, the optimality under the termination condition $\Theta_n = \delta(n-N)$ can emerge before the landscape has been fully exploited. Therefore, the optimality we observe here remain valid even if we introduce an additional condition to the termination, i.e. $\Theta_n = \delta(n-N) + \delta(S_n - K)$, such that the foraging will be terminated once the forager discovers all targets (or we assume that the forager has prior knowledge of the number of targets) and will not fall into the 'zero-gain' regime.

\subsection{The general case with $\beta>0$}
\begin{figure}[htb!]
\centering
\vspace{1in}
\includegraphics[width=6in]{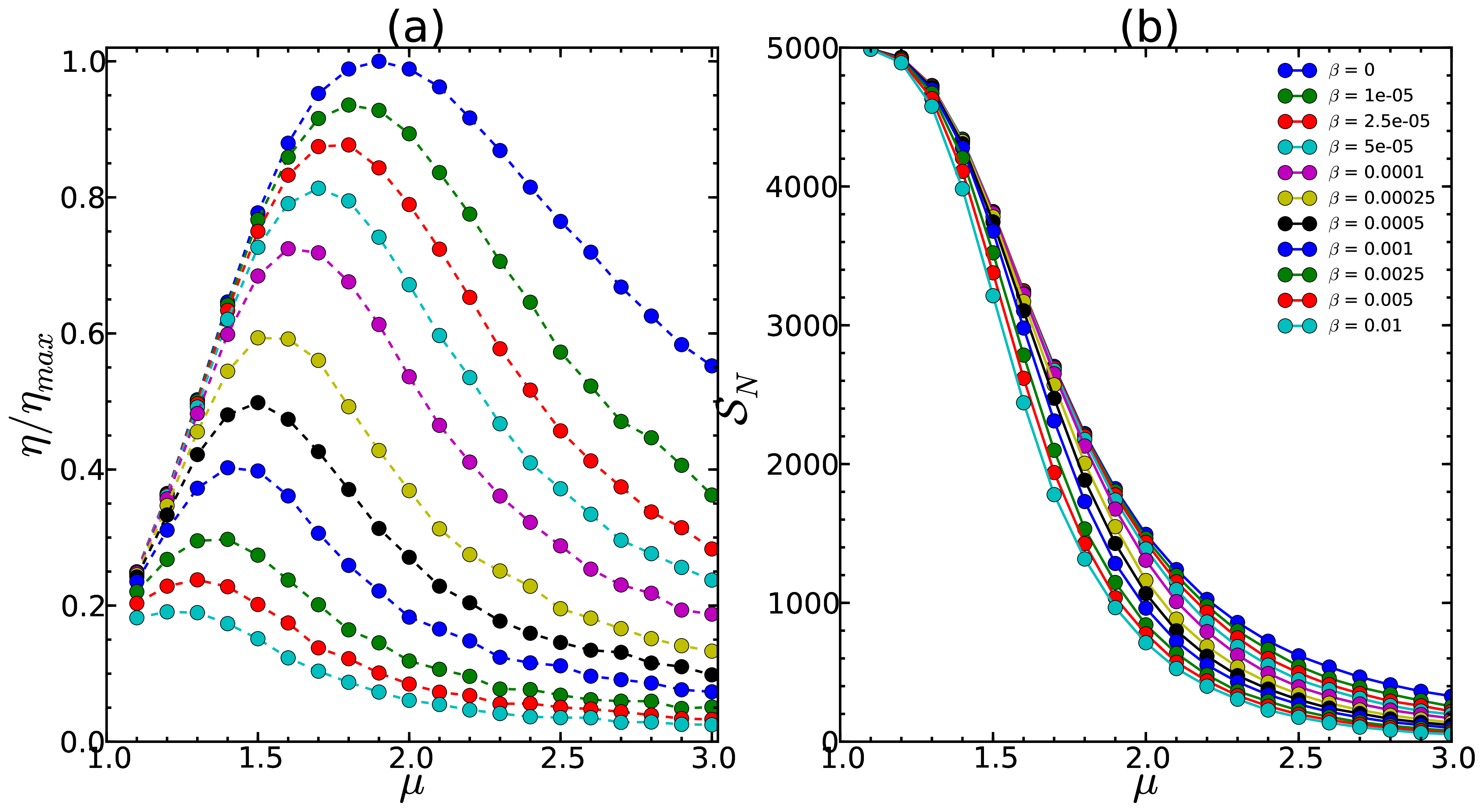}
\caption{(a) $\eta / \eta_{max}$ versus $\mu$ for various intensity of subjective returning $\beta$ from numerical simulation with the landscape size $L = 1000$, the number of patches $K = 5000$ and the termination condition $\Theta_n = \delta(n - 50000)$. The results are averaged over $100$ realizations. The dots represent simulation results and the dashed lines are a guide to the eye. (b) The total number of discovered targets $S_N$ versus $\mu$. The curves in panel (a) and (b) with different colours from top to bottom corresponding to various $\beta$ with increasing values as indicated in the legend of panel (b).  }
\label{fig5}
\end{figure}

We now turn our attention to the case with $\beta > 0$ and briefly study this case with the simple termination condition $\Theta_n = \delta(n-N)$. We perform numerical simulation with various intensity $\beta$ and exponent $\mu$, as shown in Figure \ref{fig5}(a).  As might be expected, $\beta$ has a significant impact on the search efficiency. When $\beta$ increases, the optimal exponent $\mu_{opt}$ will shift to a smaller value. 

The result here indicates that an optimal L\'{e}vy-flight search strategy can be tuned by the subjective returning which is widely observed in animals. This finding is consistent with the results in \cite{hu2011toward} that the optimal exponent which maximises the entropy of visited locations in L\'{e}vy-flight is reduced by incorporating more frequent returns in the movement. 

One should note that, the subjective returning here is a constraint for the random search. With $\mu$ and $N$ given, the search efficiency $\eta$ is a monotonically decreasing function of $\beta$, which indicates that a higher intensity of returning is harmful to the search efficiency. Such constraint can be associated with a wide range of intrinsic physical or psychological features of the forager or external environmental context of the landscape. For example, a high value of $\beta$ can represent a harsh foraging environment that drives the forager to do more frequent subjective returnings. 

We also look at the two components $S_N$ and $\mathcal{L}_N$ that gauge the search efficiency. As shown in Figure \ref{fig5}(b), with $\beta$ given, $S_N$ is a decreasing function of $\mu$  which aligns with the results for $\beta=0$ that smaller $\mu$ leading to longer steps gives rise to more discovered targets in the end. Meanwhile, with $\mu$ given, $S_N$ is a decreasing function of $\beta$. This is also intuitive since more frequent returns reduce the portion of exploration steps in the whole foraging process and the success rate of discovering new targets in each exploration step. We find that,  for a given $\beta$,  $\mathcal{L_N}$ remains a decreasing function of $\mu$, as shown in Figure \ref{fig7}(a). Similarly to the case with $\beta=0$, the emergence of the optimal exponent for exploration-return can be also explained by considering the tradeoff between increasing $S_N$ and increasing $\mathcal{L}_N$. 

\begin{figure}[htb!]
\centering
\vspace{1in}
\includegraphics[width=5.5in]{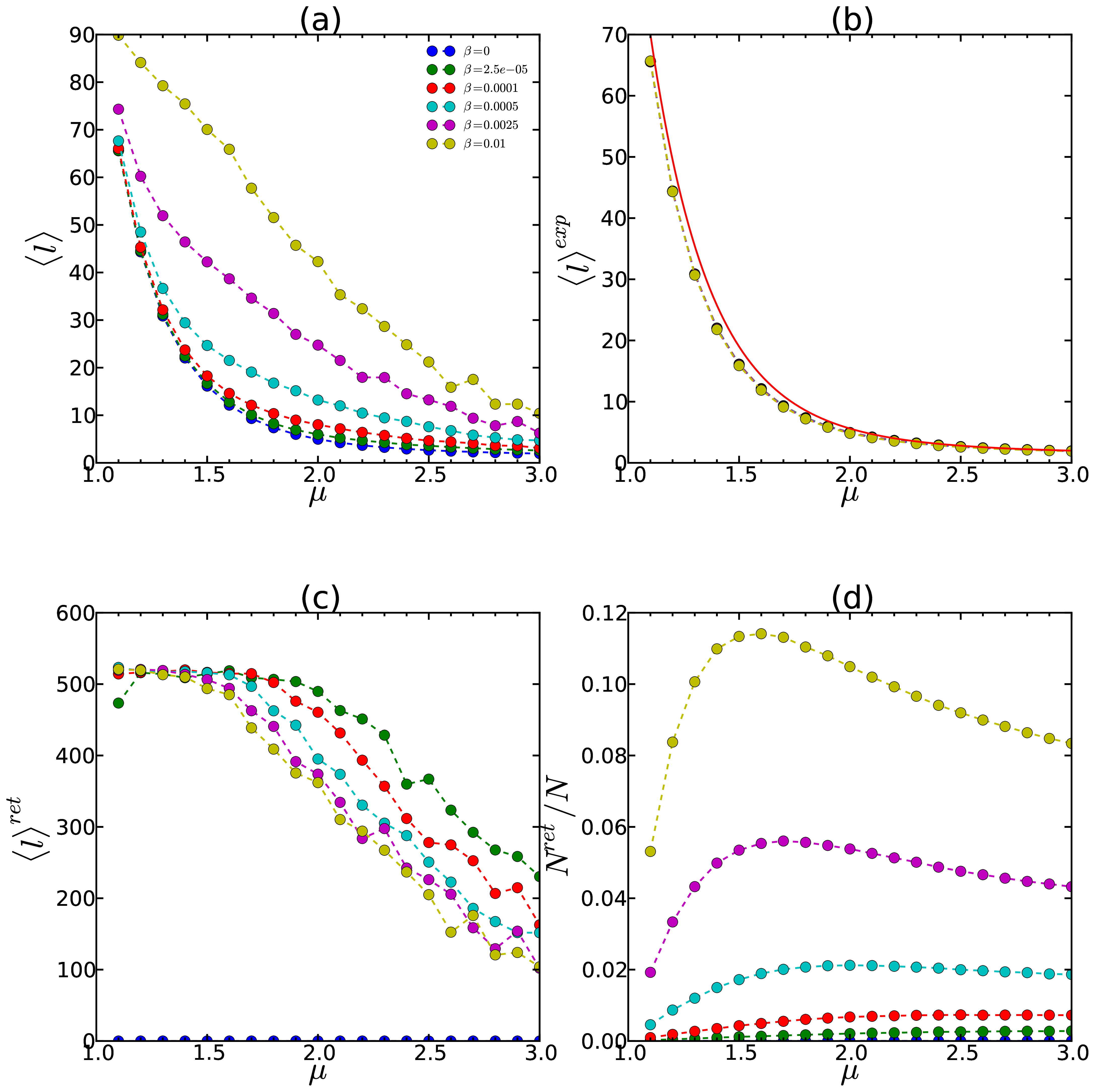}
\caption{(a) The mean step-size $\avg{l}$ versus the power-law exponent $\mu$. (b) The mean step-size in exploration $\avg{l}^{exp}$ versus $\mu$. The red solid line represents Eq.(\ref{avg_l}). (c) The mean step-size in return $\avg{l}^{ret}$ versus $\mu$. (d) The ratio of the number of return steps $N^{ret}$ to the total number of steps $N$ versus $\mu$. In panel (a) - (d), the dots represent simulations results and the dashed lines are a guide to the eye. Different colours correspond to different values of $\beta$, as indicated in the legend of panel (a). The results are obtained from numerical simulation with the landscape size $L = 1000$, the number of patches $K = 5000$ and the termination condition $\Theta_n = \delta(n - 50000)$, and are averaged over $100$ realizations. }
\label{fig6}
\end{figure}

\begin{figure}[htb!]
\centering
\vspace{1in}
\includegraphics[width=5.5in]{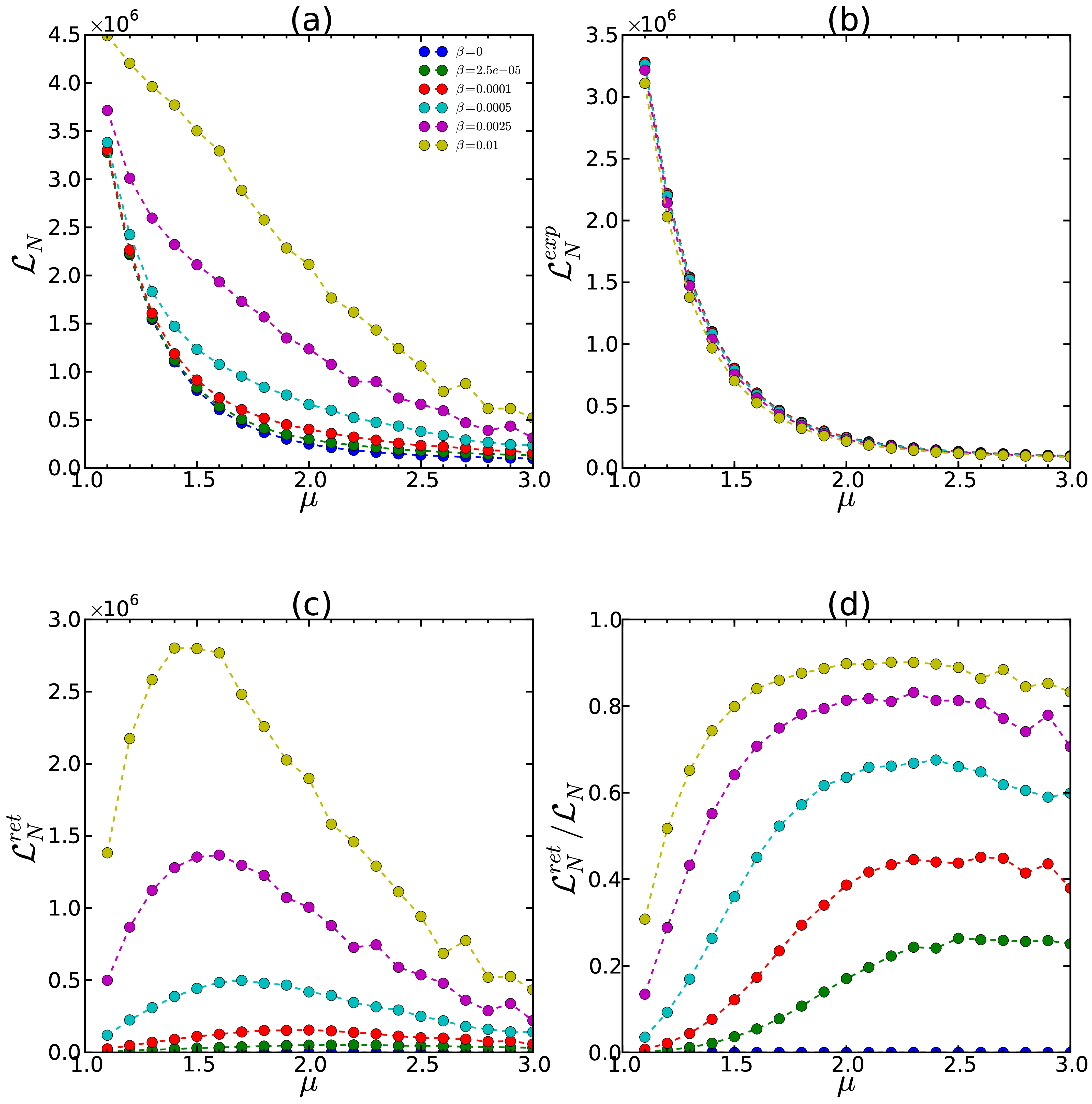}
\caption{(a) The total moving distance $\mathcal{L}_N$ versus the power-law exponent $\mu$. (b) The total moving distance in exploration $\mathcal{L}^{exp}_N$ versus $\mu$. (c) The total moving distance in return $\mathcal{L}^{ret}_N$ versus $\mu$. (d) The ratio of $\mathcal{L}^{ret}_N$ to the total number of steps $\mathcal{L}_N$ versus $\mu$. In panel (a) - (d), the dots represent simulations results and the dashed lines are a guide to the eye. Different colours correspond to different values of $\beta$, as indicated in the legend of panel (a). The results are obtained from numerical simulation with the landscape size $L = 1000$, the number of patches $K = 5000$ and the termination condition $\Theta_n = \delta(n - 50000)$, and are averaged over $100$ realizations. }
\label{fig7}
\end{figure}

Finally, we discuss some interesting characteristics regarding $\mathcal{L_N}$. In the presence of subjective returning, $\mathcal{L}_N$ is composed of two parts corresponding to two distinct phases respectively. In particular we can write $\mathcal{L}_N = \mathcal{L}_N^{exp} + \mathcal{L}_N^{ret}$ where the superscript denotes the corresponding phase. $\mathcal{L}_N$ can be also written as $\mathcal{L}_N = N \avg{l} = N^{exp} \avg{l}^{exp} + N^{ret} \avg{l}^{ret}$ where the superscripted $N$ and $\avg{l}$ denote the number of steps and mean step-size in each phase, and we have $\mathcal{L}_N^{exp} = N^{exp} \avg{l}^{exp}$, $\mathcal{L}_N^{ret} = N^{ret} \avg{l}^{ret}$ and $N = N^{exp} + N^{ret}$. The values of these quantities with respect to $\mu$ and $\beta$ are shown in Figure \ref{fig6} and Figure \ref{fig7}. We find that, for a given value of $\mu$, $\avg{l}$ and $\mathcal{L_N}$ are also a increasing function of $\beta$, which indicates high intensity of returning leads to larger mean-step size. We then find that, as expected,  $\avg{l}^{exp}$ as a function of $\mu$ (described by Eq.(\ref{avg_l})) does not change with respect to $\beta$ since the characteristics in the exploration phase is not affected by $\beta$, and with $\mu$ fixed $\mathcal{L}_N^{exp}$ is a decreasing function of $\beta$ since the portion of exploration steps decreases as $\beta$ increases. The increase of $\mathcal{L}_N$ with respect to $\beta$ is mostly contributed by the returns since $\avg{l}^{ret}$ is much larger than $\avg{l}^{exp}$. Interestingly we find that $\mathcal{L}_N^{ret} $, as well as the portion of return steps  $\mathcal{N}^{ret}/N$ , are not a monotonic function of $\mu$ and exhibit a maximum. A possible explanation for this peak is that the value of $\mu$ determines the balance between intended return and accidental rediscovery of a known target. A lower $\mu$ results in longer steps, leading to higher likelihood of the forager deciding to return. A higher $\mu$ results in shorter steps, leading the forager to revisit the same target many times consecutively by accidentally rediscovering targets (the trapping  phenomenon in the model). Following this reasoning, the peaks in $\mathcal{L}_N^{ret} $ may reflect the value of $\mu$ that maximises the aggregated return distance due to the intended and accidental returns. Fully understanding these phenomena is beyond the scope of this paper and we leave it for future study.

\section{Conclusion}

In this paper, we have presented a simple model to study L\'{e}vy-flight foraging as an exploratory process subject to termination in a finite landscape with countable targets. The first important message of our results is that, the interplay between the termination of exploiting the landscape and the environmental features such as the landscape size and resource distribution, can have an significant influence on the development of optimal foraging strategy. In particular, when the number of steps (or trips/bouts in foraging) plays a role in the termination condition, various values of $\mu_{opt}$ that captures foraging dynamics from ballistic motion to L\'{e}vy-flight to Brownian motion, can evolve in response to other constrained factors such as the landscape size, the number of targets as well as the intensity of subjective returning. 

It would be interesting for researchers to test our model in empirical data. One direct observable consequence suggested by our study is that, the value of $\mu_{opt}$ may have a relationship with some dependent quantities such as the total moving distance (energy expenditure) or the number of steps (trips/bouts) of a foraging process, as well as the size or the resource distribution of the foraging area. This dependence may be tested by measuring the correlation between the value of $\mu_{opt}$ and these quantities among individuals.  

This idealised model allows a variety of future extensions. In the current approach, we do not consider energy intake and consumption, and simply assume that the forager always has enough energy to perform long flights. One important improvement to the model is to incorporate an adaptive foraging strategy based on available energy. As a supplement to this work, one can also study other termination conditions to capture more complex internal regularity, memory mechanisms or decision-making of foraging \cite{perez2011synergy, janson2007experimental, cheng1998mechanisms, volchenkov2013exploration, dukas1998cognitive}. Moreover, one can consider time-variant and heterogenous distribution of targets in the landscape, which can lead to a study of how the foraging strategy adapts to the environmental changes in the presence of cognition and memory. It is also worth extending the model to incorporate multiple foragers and studying the collective process of competition and cooperation in foraging \cite{giraldeau2000social}. Finally, although it is initially proposed to study animal foraging,  our model can be generally used in other random search or biological encountering processes as well.

\section*{Acknowledgements}

This research was supported by a grant from CSIRO's Sensor and Sensor Networks Transformational Capability Platform, CSIRO's OCE Postdoctoral Scheme, the Australian Government Department of the Environment's the National Environmental Research Program and the Rural Industries Research Development Corporation (through funding from the Commonwealth of Australia, the State of New South Wales and the State of Queensland under the National Hendra Virus Research Program).


\begin{thebibliography}{10}

\bibitem{smouse2010stochastic} 
Smouse PE, Focardi S, Moorcroft PR, Kie JG, Forester JD, Morales JM. 
\newblock 2010 \R{Stochastic modelling of animal movement}.
\newblock \textit{Philos. T. R. Soc. B} \textbf{365}, 2201--2211.

\bibitem{bartumeus2007levy} 
Bartumeus F.
\newblock 2007 \R{L{\'e}vy processes in animal movement: an evolutionary hypothesis}.
\newblock \textit{Fractals} \textbf{15}, 151--162.

\bibitem{ortiz2006use} 
Ortiz-Pelaez A, Pfeiffer D, Soares-Magalhaes R, Guitian F.
\newblock 2006 \R{Use of social network analysis to characterize the pattern of animal
  movements in the initial phases of the 2001 foot and mouth disease (FMD)
  epidemic in the UK}.
\newblock \textit{Prev. Vet. Med.} \textbf{76}, 40--55.

\bibitem{caro1999behaviour} 
Caro T.
\newblock 2009 \R{The behaviour--conservation interface}.
\newblock \textit{Trends Ecol. Evol.} \textbf{14}, 366--369.

\bibitem{bousquet2004multi} 
Bousquet F, Le~Page C.
\newblock 2004 \R{Multi-agent simulations and ecosystem management: a review}.
\newblock \textit{Ecol. Model.} \textbf{176}, 313--332.

\bibitem{viswanathan1996levy} 
Viswanathan G, Afanasyevt V.
\newblock 1996 \R{L{\'e}vy flight search patterns of wandering albatrosses}.
\newblock \textit{Nature} \textbf{381}, 30.

\bibitem{humphries2012foraging} 
Humphries NE, Weimerskirch H, Queiroz N, Southall EJ, Sims DW.
\newblock 2012 \R{Foraging success of biological L{\'e}vy flights recorded in situ.}
\newblock \textit{Proc. Natl Acad. Sci.} \textbf{109}, 7169--7174.

\bibitem{ramos2004levy} 
Ramos-Fernandez G, Mateos JL, Miramontes O, Cocho G, Larralde H, Ayala-Orozco
  B.
\newblock 2004 \R{L{\'e}vy walk patterns in the foraging movements of spider monkeys
  (Ateles geoffroyi)}.
\newblock \textit{Behav. Ecol. Sociobiol.} \textbf{55}, 223--230.

\bibitem{reynolds2007displaced} 
Reynolds AM, Smith AD, Menzel R, Greggers U, Reynolds DR, Riley JR.
\newblock 2007 \R{Displaced honey bees perform optimal scale-free search flights}.
\newblock \textit{Ecology} \textbf{88}, 1955--1961.

\bibitem{focardi2009adaptive} 
Focardi S, Montanaro P, Pecchioli E.
\newblock 2009 \R{Adaptive L{\'e}vy walks in foraging fallow deer}.
\newblock \textit{PloS ONE} \textbf{4}, e6587.

\bibitem{sims2008scaling} 
Sims DW, Southall EJ, Humphries NE, Hays GC, Bradshaw CJ, Pitchford JW, et~al.
\newblock 2008 \R{Scaling laws of marine predator search behaviour}.
\newblock \textit{Nature} \textbf{451}, 1098--1102.

\bibitem{hills2013adaptive} 
Hills TT, Kalff C, Wiener JM.
\newblock 2013 \R{Adaptive L{\'e}vy processes and area-restricted search in human
  foraging}.
\newblock \textit{PloS ONE} \textbf{8}, e60488.


\bibitem{raichlen2014evidence} 
Raichlen DA, Wood BM, Gordon AD, Mabulla AZ, Marlowe FW, Pontzer H.
\newblock 2014 \R{Evidence of L{\'e}vy walk foraging patterns in human
  hunter--gatherers.}
\newblock \textit{Proc. Natl Acad. Sci.} \textbf{111}, 728--733.

\bibitem{klafter2011first} 
Klafter J, Sokolov IM.
\newblock 2011 \textit{First steps in random walks: from tools to applications}.
\newblock Oxford University Press.

\bibitem{benhamou2007many} 
Benhamou S.
\newblock 2007 \R{How many animals really do the Levy walk?}
\newblock \textit{Ecology} \textbf{88}, 1962--1969.

\bibitem{edwards2007revisiting} 
Edwards AM, Phillips RA, Watkins NW, Freeman MP, Murphy EJ, Afanasyev V, Buldyrev SV, da Luz MG, Raposo EP, Stanley HE, et~al.
\newblock 2007 \R{Revisiting L{\'e}vy flight search patterns of wandering albatrosses,
  bumblebees and deer.}
\newblock \textit{Nature}, \textbf{449}, 1044--1048.

\bibitem{edwards2011overturning} 
Edwards AM.
\newblock 2011 \R{Overturning conclusions of L{\'e}vy flight movement patterns by
  fishing boats and foraging animals.}
\newblock \textit{Ecology} \textbf{92}, 1247--1257.



\bibitem{viswanathan2008levy} 
Viswanathan G, Raposo E, Da~Luz M.
\newblock 2008 \R{L{\'e}vy flights and superdiffusion in the context of biological
  encounters and random searches.}
\newblock \textit{Phys. Life Rev.} \textbf{5}, 133--150.

\bibitem{viswanathan2011physics} 
Viswanathan GM.
\newblock 2011 \textit{The physics of foraging: an introduction to random searches and
  biological encounters.}
\newblock Cambridge University Press.

\bibitem{james2011assessing} 
James A, Plank MJ, Edwards AM.
\newblock 2011 \R{Assessing L{\'e}vy walks as models of animal foraging.}
\newblock \textit{J. R. Soc. Interface} \textbf{8}, 1233--1247.

\bibitem{bartumeus2009optimal} 
Bartumeus F, Catalan J.
\newblock 2009 \R{Optimal search behavior and classic foraging theory.}
\newblock \textit{J. Phys. A } \textbf{42}, 434002.

\bibitem{viswanathan2000levy} 
Viswanathan G, Afanasyev V, Buldyrev SV, Havlin S, Da~Luz M, Raposo E, et~al.
\newblock 2000 \R{L{\'e}vy flights in random searches.}
\newblock \textit{Physica A} \textbf{282}, 1--12.

\bibitem{bartumeus2013stochastic} 
Bartumeus F, Raposo EP, Viswanathan GM, da~Luz MG.
\newblock 2013 \R{Stochastic optimal foraging theory.}
\newblock In \textit{Dispersal, Individual Movement and Spatial Ecology},   pp. 3--32. Springer.

\bibitem{bartumeus2005animal} 
Bartumeus F, da~Luz MGE, Viswanathan G, Catalan J.
\newblock 2005 \R{Animal search strategies: a quantitative random-walk analysis.}
\newblock \textit{Ecology} \textbf{86}, 3078--3087.

\bibitem{da2009random} 
da~Luz MG, Grosberg A, Raposo EP, Viswanathan GM.
\newblock 2009 \R{The random search problem: trends and perspectives.}
\newblock \textit{J. Phys. A} \textbf{42}, 430301.

\bibitem{reynolds2009levy} 
Reynolds A, Rhodes C.
\newblock 2009 \R{The L{\'e}vy flight paradigm: random search patterns and mechanisms.}
\newblock \textit{Ecology} \textbf{90}, 877--887.

\bibitem{viswanathan1999optimizing} 
Viswanathan G, Buldyrev SV, Havlin S, Da~Luz M, Raposo E, Stanley HE.
\newblock 1999 \R{Optimizing the success of random searches}.
\newblock \textit{Nature} \textbf{401}, 911--914.


\bibitem{benichou2007minimal} 
Benichou O, Loverdo C, Moreau M, Voituriez R.
\newblock 2007 \R{A minimal model of intermittent search in dimension two.}
\newblock \textit{J. Phys. } \textbf{19}, 065141.


\bibitem{gonzalez2008understanding} 
Gonzalez MC, Hidalgo CA, Barabasi AL.
\newblock 2008 \R{Understanding individual human mobility patterns.}
\newblock \textit{Nature} \textbf{453}, 779--782.

\bibitem{hays2012high} 
Hays GC, Bastian T, Doyle TK, Fossette S, Gleiss AC, Gravenor MB, et~al.
\newblock 2012 \R{High activity and L{\'e}vy searches: jellyfish can search the water
  column like fish.}
\newblock \textit{Proc. R. Soc. B} \textbf{279}, 465--473.

\bibitem{santos2004optimal} 
Santos M, Raposo E, Viswanathan G, Da~Luz M.
\newblock 2004 \R{Optimal random searches of revisitable targets: crossover from
  superdiffusive to ballistic random walks.}
\newblock \textit{Europhys. Lett.} \textbf{67}, 734.

\bibitem{raposo2011landscape} 
Raposo E, Bartumeus F, Da~Luz M, Ribeiro-Neto P, Souza T, Viswanathan G.
\newblock 2011 \R{How landscape heterogeneity frames optimal diffusivity in searching
  processes.}
\newblock \textit{PLoS Comput. Biol.} \textbf{7}, e1002233.

\bibitem{palyulin2014levy} 
Palyulin VV, Chechkin AV, Metzler R.
\newblock 2014 \R{L{\'e}vy flights do not always optimize random blind search for
  sparse targets.}
\newblock \textit{Proc. Natl Acad. Sci.} \textbf{111}, 2931--2936.


\bibitem{song2010modelling} 
Song C, Koren T, Wang P, Barab{\'a}si AL.
\newblock 2010 \R{Modelling the scaling properties of human mobility.}
\newblock \textit{Nat. Phys.} \textbf{6}, 818--823.

\bibitem{niv2002evolution} 
Niv Y, Joel D, Meilijson I, Ruppin E.
\newblock 2002 \R{Evolution of reinforcement learning in uncertain environments: A
  simple explanation for complex foraging behaviors.}
\newblock \textit{Adapt. Behav.} \textbf{10}, 5--24.

\bibitem{zhao2011social} 
Zhao K, Stehl{\'e} J, Bianconi G, Barrat A.
\newblock 2011 \R{Social network dynamics of face-to-face interactions.}
\newblock \textit{Phys. Rev. E} \textbf{83}, 056109.

\bibitem{reynolds2008optimal} 
Reynolds A.
\newblock 2008 \R{Optimal random L{\'e}vy-loop searching: New insights into the
  searching behaviours of central-place foragers.}
\newblock \textit{Europhys. Lett.} \textbf{82}, 20001.


\bibitem{steingrimsson2008multiple} 
Steingr{\'\i}msson S{\'O}, Grant JW.
\newblock 2008 \R{Multiple central-place territories in wild young-of-the-year Atlantic
  salmon Salmo salar?}
\newblock \textit{J. Anim Ecol.} \textbf{77}, 448--457.

\bibitem{buldyrev2001average} 
Buldyrev S, Havlin S, Kazakov AY, da~Luz M, Raposo E, Stanley H, et~al.
\newblock 2001 \R{Average time spent by L{\'e}vy flights and walks on an interval with
  absorbing boundaries.}
\newblock \textit{Phys. Rev. E} \textbf{64}, 041108.

\bibitem{hu2011toward} 
Hu Y, Zhang J, Huan D, Di Z.
\newblock 2011 \R{Toward a general understanding of the scaling laws in human and
  animal mobility.}
\newblock \textit{Europhys. Lett.} \textbf{96}, 38006.

\bibitem{perez2011synergy} 
P{\'e}rez-Reche FJ, Ludlam JJ, Taraskin SN, Gilligan CA.
\newblock 2011 \R{Synergy in spreading processes: from exploitative to explorative
  foraging strategies.}
\newblock \textit{Phys. Rev. Lett.} \textbf{106}, 218701.

\bibitem{janson2007experimental} 
Janson CH.
\newblock 2007 \R{Experimental evidence for route integration and strategic planning in
  wild capuchin monkeys.}
\newblock \textit{Anim. Cogn.} \textbf{10}, 341--356.

\bibitem{cheng1998mechanisms} 
Cheng K, Spetch ML.
\newblock 1998 \textit{Mechanisms of landmark use in mammals and birds.}
\newblock Oxford University Press.

\bibitem{volchenkov2013exploration} 
Volchenkov D, Helbach J, Tscherepanow M, K{\"u}hnel S.
\newblock 2013 \R{Exploration--exploitation trade-off features a saltatory search
  behaviour.}
\newblock \textit{J. R. Soc. Interface} \textbf{10}.

\bibitem{dukas1998cognitive} 
Dukas R.
\newblock 1998 \textit{Cognitive ecology: the evolutionary ecology of information processing
  and decision making.}
\newblock University of Chicago Press.

\bibitem{giraldeau2000social} 
Giraldeau LA, Caraco T.
\newblock 2000 \textit{Social foraging theory.}
\newblock Princeton University Press.





\end{thebibliography}

\clearpage

\clearpage

\clearpage

\clearpage

\clearpage

\clearpage

\clearpage

\clearpage

\end{document}